# *Ensemble-Guided Distillation for Compact and Robust Acoustic Scene Classification on Edge Devices*

Hossein Sharifi
Electrical Engineering Department
*Sharif University of Technology*
Tehran, Iran
sharifir.hossein@ee.sharif.edu

Behnam Raoufi
Electrical Engineering Department
*Sharif University of Technology*
Tehran, Iran
behnam.raoufi93@sharif.edu

Mahdy Ramezani
Electrical Engineering Department
*Sharif University of Technology*
Tehran, Iran
ramezani.mm@ee.sharif.edu

Khosrow Hajsadeghi
Electrical Engineering Department
Sharif University of Technology
Tehran, Iran
ksadeghi@sharif.edu

Saeed Bagheri Shouraki
Electrical Engineering Department
*Sharif University of Technology*
Tehran, Iran
bagheri-s@sharif.edu

***Abstract*— We present a compact, quantization-ready acoustic scene classification (ASC) framework that couples an efficient student network with a learned teacher ensemble and knowledge distillation. The student backbone uses stacked depthwise-separable "expand–depthwise–project" blocks with global response normalization to stabilize training and improve robustness to device and noise variability, while a global pooling head yields class logits for efficient edge inference. To inject richer inductive bias, we assemble a diverse set of teacher models and learn two complementary fusion heads: z1, which predicts per-teacher mixture weights using a student-style backbone, and z2, a lightweight MLP that performs per-class logit fusion. The student is distilled from the ensemble via temperature-scaled soft targets combined with hard labels, enabling it to approximate the ensemble's decision geometry with a single compact model. Evaluated on the TAU Urban Acoustic Scenes 2022 Mobile benchmark, our approach achieves state-of-the-art (SOTA) results on the TAU dataset under matched edge-deployment constraints, demonstrating strong performance and practicality for mobile ASC.***

## I. INTRODUCTION

Acoustic Scene Classification (ASC) aims to identify the environment depicted in an audio recording—such as airport, metro station, or park—and has become a core capability for context-aware mobile and embedded devices. Despite substantial progress driven by large-scale pretraining and high-capacity neural architectures [1], real-world deployment remains constrained by tight latency, memory, and energy budgets on edge hardware [2]. Compounding these constraints is a persistent device-shift problem: models trained on audio from specific microphones or smartphones often degrade when confronted with recordings from unseen or mismatched devices [3].

This work addresses both efficiency and robustness with a compact, quantization-ready student network distilled from a diverse teacher ensemble. The student backbone is built from depthwise-separable "expand–depthwise–project" blocks and incorporates Global Response Normalization (GRN) to stabilize optimization in small models and to improve tolerance to noise and device variability [4]. To transfer rich inductive bias without sacrificing deployability, we construct an ensemble of heterogeneous teachers (including stronger variants of the student as well as complementary architectures) and learn two lightweight fusion heads: z1, which predicts per-teacher mixture weights using a student-style backbone, and z2, a minimal MLP that performs per-class logit fusion. Knowledge distillation then encourages the student to align with the ensemble's softened predictions while respecting ground-truth labels, enabling a single compact model to approximate the decision geometry of many [5].

We evaluate on TAU Urban Acoustic Scenes 2022 Mobile (TAU-UAS 2022 Mobile), a benchmark explicitly designed to reflect mobile/edge conditions and multi-device capture [6]. To strengthen generalization, we leverage broad pretraining on CochlScene, AudioSet, and ESC-50 [7] [8] [9] before adapting to TAU-UAS 2022 Mobile, and we incorporate impulse-response–based augmentation to simulate device characteristics [10]. Under matched edge-deployment constraints typical of DCASE systems, our approach achieves state-of-the-art (SOTA) results on the TAU dataset, demonstrating that careful combination of compact architectures, ensemble-guided learning, and device-aware training can close the gap between research-grade accuracy and practical on-device inference.

## II. RELATED WORK

Acoustic Scene Classification (ASC) has progressed substantially over the past decade, driven by the availability of large datasets and increasingly powerful deep learning architectures. Early ASC systems primarily leveraged hand-crafted spectral or statistical features combined with conventional classifiers, but deep convolutional networks rapidly became dominant following the introduction of large-scale transfer learning resources such as AudioSet [11]. Subsequent research has increasingly focused on balancing accuracy with the practical constraints of mobile and embedded deployment. Compact convolutional models based on depthwise-separable operations, inspired by MobileNetV2's inverted residual blocks [12], have been widely adopted in embedded ASC due to their favorable accuracy-efficiency tradeoff. These architectures have been further enhanced through improved normalization mechanisms such as Global Response Normalization (GRN), introduced by Pan et al. for stabilizing training and improving robustness in compact models. More recent systems, including PaSST [13] and lightweight ResNet variants [14], have demonstrated strong results by combining efficient attention mechanisms or optimized residual structures with spectrogram-based front ends.

Despite architectural advances, ASC performance continues to be hindered by domain and device mismatch, a long-standing challenge highlighted in DCASE evaluations [15]. Approaches addressing this challenge include multi-device training, adversarial domain alignment, impulse-response augmentation, and metadata-aware conditioning [16] [17]. Large-scale pretraining on heterogeneous audio corpora such as AudioSet , ESC-50 , and CochlScene has also proven effective for mitigating device shift by improving the generality of learned representations.

Knowledge distillation (KD), introduced by Hinton et al. [18] , has become a standard strategy for compressing high-capacity audio models into deployable student networks while retaining competitive accuracy. Extensions to KD for audio classification include multi-teacher distillation , logit-level smoothing under class imbalance, and temperature-aware fusion strategies. Ensemble distillation, wherein a student approximates the collective decision geometry of several heterogeneous teachers, has shown particular promise for ASC due to the complementary inductive biases of diverse models.

Learned ensemble fusion mechanisms further enhance the utility of teacher models. Mixture-of-experts–style weighting networks [19] and lightweight per-class fusion layers have been explored for audio tasks to improve robustness in complex acoustic conditions. Such approaches outperform static averaging by allowing sample-dependent aggregation of teacher outputs. In ASC specifically, ensemble-based systems have demonstrated state-of-the-art performance in recent DCASE challenges under matched edge constraints [20].

## III. APPROACHES

This section details the student model, the construction of the network input, the teacher–ensemble setup, and the knowledge-distillation procedure.

### *A. Student Model Architecture*

As shown in Fig. 1, downstream of the learnable front end, the backbone is a compact, quantization-friendly CNN composed of stacked, depthwise-separable ‘expand–depthwise–project’ blocks, with lightweight residual connections where tensor shapes align. A Global Response Normalization module after each block rescales activations based on global energy, stabilizing training in the small-model regime and improving robustness to device and noise variation. Blocks are organized into three successive stages with modest width growth aligned to hardware-friendly channel granularities; selective downsampling balances receptive field with latency. A linear classification head maps the final features directly to class logits via global average pooling. The network is prepared for quantization-aware training and exposes operator fusion to collapse convolution– normalization–activation sequences for efficient integer inference. In our student configuration the model contains about 60k parameters and requires roughly 30M MACs per inference.

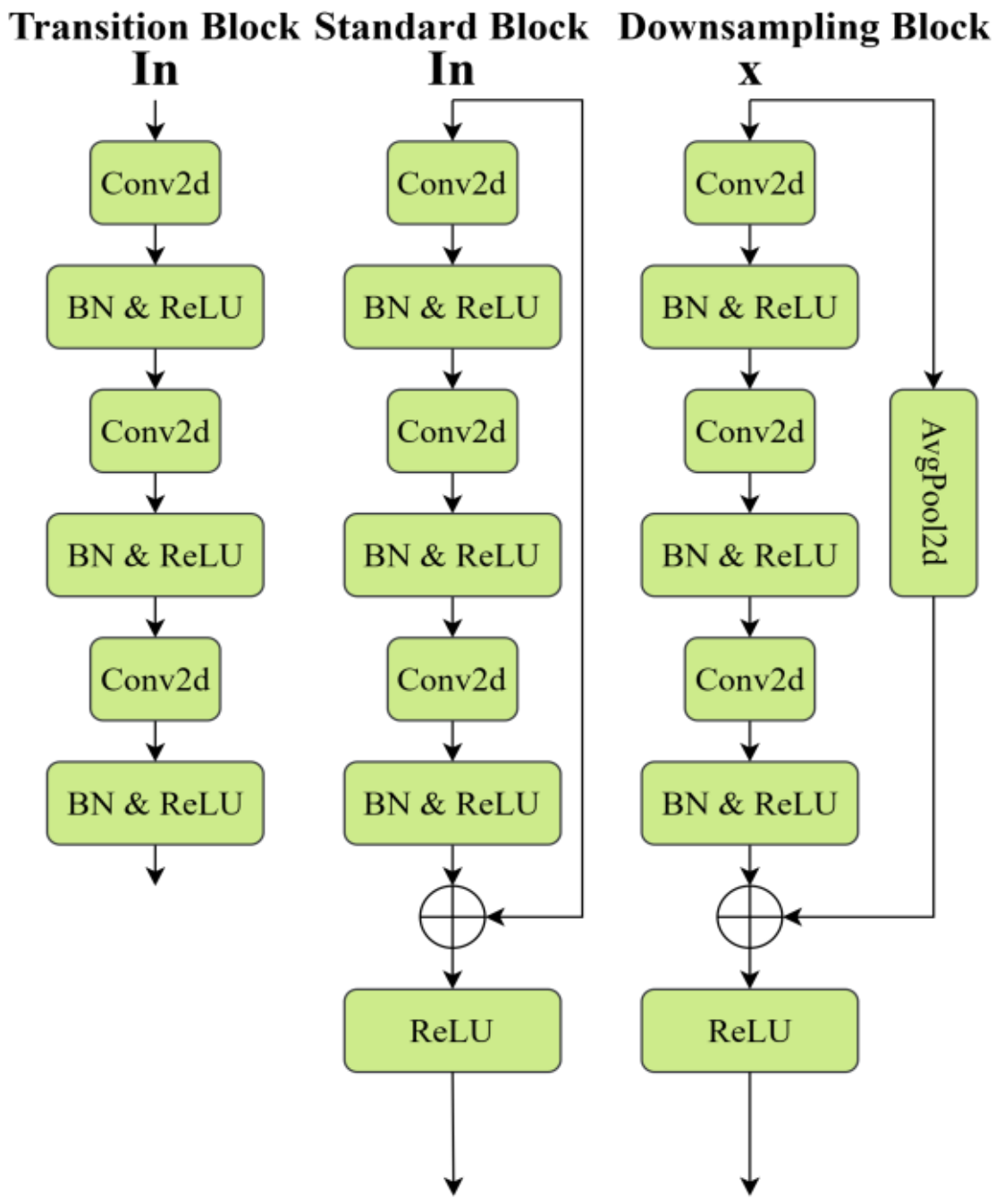

Fig. 1. Student network architecture

### *B. Ensembling teacher models*

We trained an eight-member teacher pool and two lightweight combiner networks to learn per-sample optimal ensembling. For completeness, we note that this configuration was chosen simply to maintain consistency across all experimental phases without introducing any architectural deviations. We incorporated the previously pre-trained teacher structures into the z1 and z2 framework so that all components could be trained jointly and refined together, after which the teacher set—comprising five higher-capacity variants of our student architecture (each trained with different hyperparameters to encourage diversity), one ResNet, one PaSST, and BEATs—was used while we learned two complementary ensemble functions, z1 and z2. A key motivation for using the z1 and z2 combiners is that they remove the need to train teachers on separate datasets, enabling joint training while still producing meaningful ensemble coefficients. Based on Fig. 2 the z1 network reuses the exact student backbone and differs only in its final layer, which is replaced to produce one score per teacher; after a softmax, these scores act as mixture coefficients that weight the teachers’ logits, yielding a sample-specific fused prediction. The z2 network is a minimal MLP with a single hidden layer that takes the teachers’ outputs as input and maps them directly to class logits, learning nonlinear combinations that are not restricted to convex weighting. We trained z1 and z2 jointly with the teachers using a composite loss: z1 is supervised via cross-entropy on the weighted teacher mixture, and z2 is supervised directly on its own logits; the final objective balances these terms so that both the mixture weights and the direct combiner improve in tandem. After this joint phase, all teachers were fine-tuned on the main TAU Urban dataset to align their decision boundaries with the target distribution. The effectiveness of this strategy—eight diverse teachers, a student-backbone combiner (z1), and a lightweight MLP combiner (z2)—is reflected in the Results section.

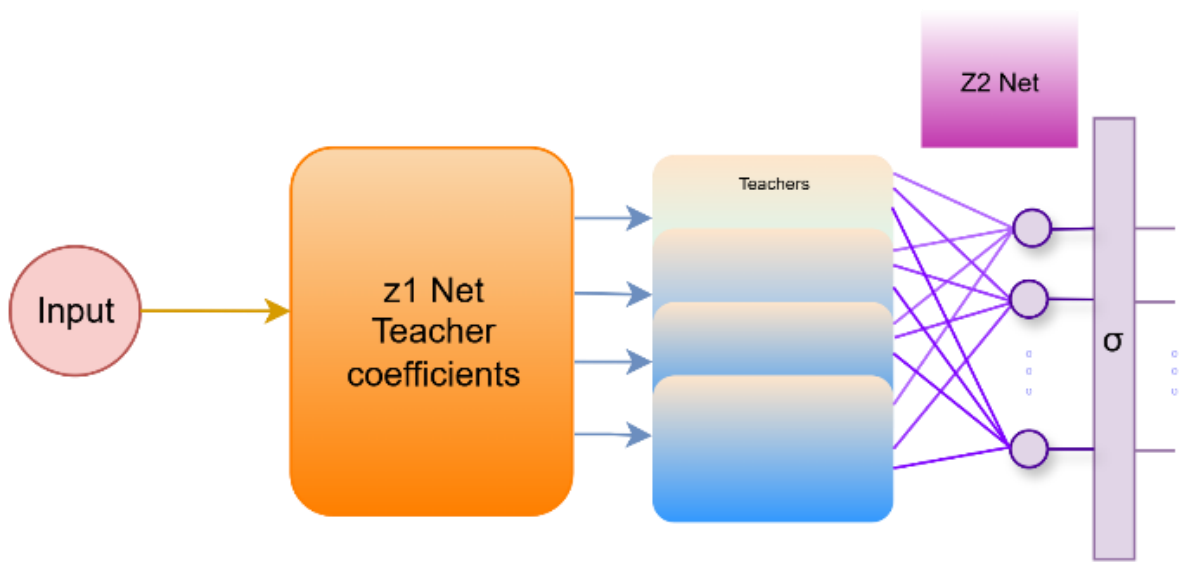

Fig. 2. Teacher-ensemble architecture

### C. *Knowledge Distillation*

Knowledge distillation transfers the behavior of one or more accurate teacher models to a compact student by complementing hard-label supervision with soft targets that capture inter-class relations . Based on Fig. 3 Given an input–label pair (x,y), train the student to fit the ground-truth label via cross-entropy and; match the teacher's softened distribution via a temperature-scaled KL term, combined with a tunable weight to balance both objectives.

$$q_T(x) = softmax\left(\frac{t_{ens}(x)}{T}\right), p_T(x) = softmax(z(x)\backslash T) \quad (1)$$

$$L_{CE}(x,y) = -\log\left(softmax\big(z(x)\big)_y\right) \quad (2)$$

$$L_{KD}(x) = T^2 KL(q_T(x)||p_T)$$

$$L_{KD}(x) = T^2 \sum_{c=1}^{C} q_T^{(c)}(x)[log q_T^c(x) - \log\,(P_T^{(c)}(x))] \quad (3)$$

$$L(x,y) = (1-\alpha)L_{CE}(x,y) + \alpha L_{KD}(x) \quad (4)$$

In (1), $t_ens(x)$ is the ensemble teacher output introduced in the previous section; applying a temperature T > 0 before the softmax yields softened teacher probabilities $q_T(x)$. The student logits z(x) are softened analogously to $p_T(x)$. In (2), $L_{CE}$ is the standard hard-label cross-entropy between the student's untempered distribution and the ground-truth class y. In (3), $L_{KD}$ encourages the student to match the teacher's softened distribution via the Kullback–Leibler divergence; the factor $T^2$ preserves gradient magnitudes across temperatures, a conventional choice in KD. In (4), $\alpha \in [0, 1]$ trades off the two signals: $\alpha = 0$ reduces to pure cross-entropy, while $\alpha = 1$ relies entirely on distillation. Here, C is the number of classes, $q_T^c\ (x)$ and $p_T^c(x)$ denote the c-th class probabilities of the teacher and student at temperature T.In practice only z(x) receives gradients from (4), allowing the student to approximate the ensemble's decision geometry while respecting the ground-truth labels.

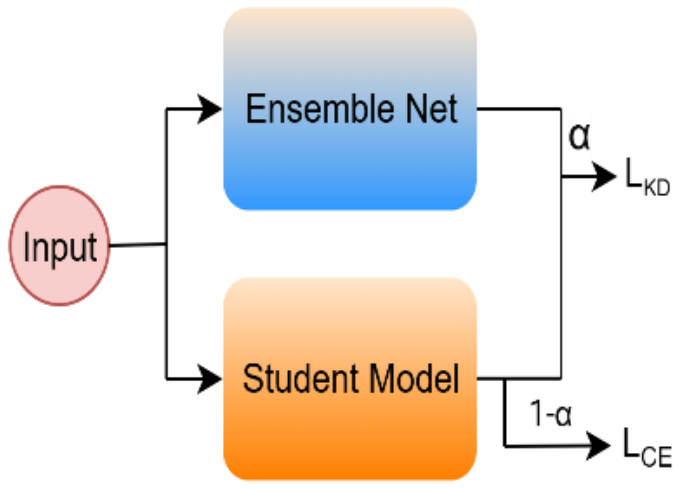

Fig. 3. Illustration of knowledge distillation with teacher ensemble network. In this structure, only the student network is trainable.

## IV. Methodology and Experiments

### A. *Datasets*

1)TAU Urban Acoustic Scenes 2022 Mobile (TAU-UAS 2022 Mobile):

We adopt the TAU Urban Acoustic Scenes 2022 Mobile dataset as our primary evaluation benchmark. The corpus contains 1-second audio segments recorded across 10 urban acoustic scenes (e.g., airport, shopping mall, metro station, pedestrian street, public square, traffic street, tram, bus, underground metro, and park), captured simultaneously with multiple handheld devices to reflect mobile/edge conditions. Our goal is to improve accuracy on this dataset and to compare against recent models specifically designed for edge devices under matched model size and compute (identical parameter counts and multiply–accumulate operations). Unless noted otherwise, all model selection and ablations are reported on TAU-UAS 2022 Mobile.

2) CochlScene:

Cochl Acoustic Scene Dataset (CochlScene) is an acoustic scene classification corpus collected entirely via crowdsourcing, following structured acquisition guidelines and quality control to ensure diverse real-world recording conditions. We use CochlScene for large-scale pre-training of our models prior to fine-tuning on TAU-UAS 2022 Mobile.

3) AudioSet:

AudioSet is a large-scale, weakly labeled collection of ~2 million 10-second audio clips from YouTube spanning an ontology of 527 sound event classes. We leverage AudioSet for broad acoustic pre-training to improve representation learning before task-specific adaptation on TAU-UAS 2022 Mobile.

4) ESC-50:

ESC-50 is a curated benchmark of 2,000 five-second environmental recordings organized into 50 classes (40 clips per class) with five standard cross-validation folds. We include ESC-50 in pre-training to regularize the models with clean, balanced examples, and we use it for auxiliary validation during development.

5) IEEE ICME 2024 Acoustic Scene Classification (ICME ASC 2024):

Released for the ICME 2024 Grand Challenge on Semi-supervised Acoustic Scene Classification under Domain Shift. The dataset is derived from the Chinese Acoustic Scene (CAS-2023) collection and covers 10 everyday scenes with >130 hours of 10-s clips recorded in 2023 across 22 Chinese cities using three industrial devices; the development set provides ~24 hours with a small labeled portion (~4 h) and a large unlabeled portion (~20 h) to facilitate semi-supervised learning under geographic/device shift. Systems are evaluated by macro-average accuracy, and we follow the official protocol and nomenclature in our experiments.

### B. *Training Configuration*

All networks are optimized using the Adam optimizer with a learning rate of $1 \times 10-4$ and a batch size of 64 for a total of 80 epochs. To improve robustness and enable better generalization across heterogeneous recording hardware, we incorporate impulse response (IR) augmentation to generate device-shifted signal variants. This augmentation process is particularly emphasized for the main dataset, where synthetic transformations are applied to emulate acoustics representative of unseen or mismatched recording devices. Based on Fig. 4, IR augmentation is performed dynamically during training. In this scheme, the degree of augmentation is adaptively modulated according to the energy characteristics of each input signal. Signals exhibiting higher energy levels undergo proportionally stronger augmentation, ensuring that the transformation intensity remains consistent with the underlying acoustic content. This adaptive strategy both diversifies the training space and reduces overfitting to specific device signatures, ultimately improving cross-device generalization performance.

## V. RESULTS AND ANALYSIS

### A. *Teacher Ensemble Results (z1, z2)*

In ensemble network the z1 head receives a mel-spectrogram input and outputs T logits—one score per teacher model—where T is the number of teachers. After a softmax, these scores become sample-adaptive mixture coefficients (dimension T). The fused prediction is then computed as a weighted sum of the T teacher logits over the target classes. The z2 head performs per-class fusion: for each class, it stacks the T teacher logits for that class and applies a small class-specific linear layer to produce the final logit; concatenating all classes forms the output vector. Both heads are trained with cross-entropy: z1 on the fused logits and z2 on its per-class outputs. Based on Table I, we have 8 teacher models, 5 of which are derivatives of the student model, and the other 3 are: ResNet, one PaSST, and BEATs. Furthermore, based on the table and the obtained results, for training the student, we use the combination of the outputs of the z1 network and the averaged output of the teachers.

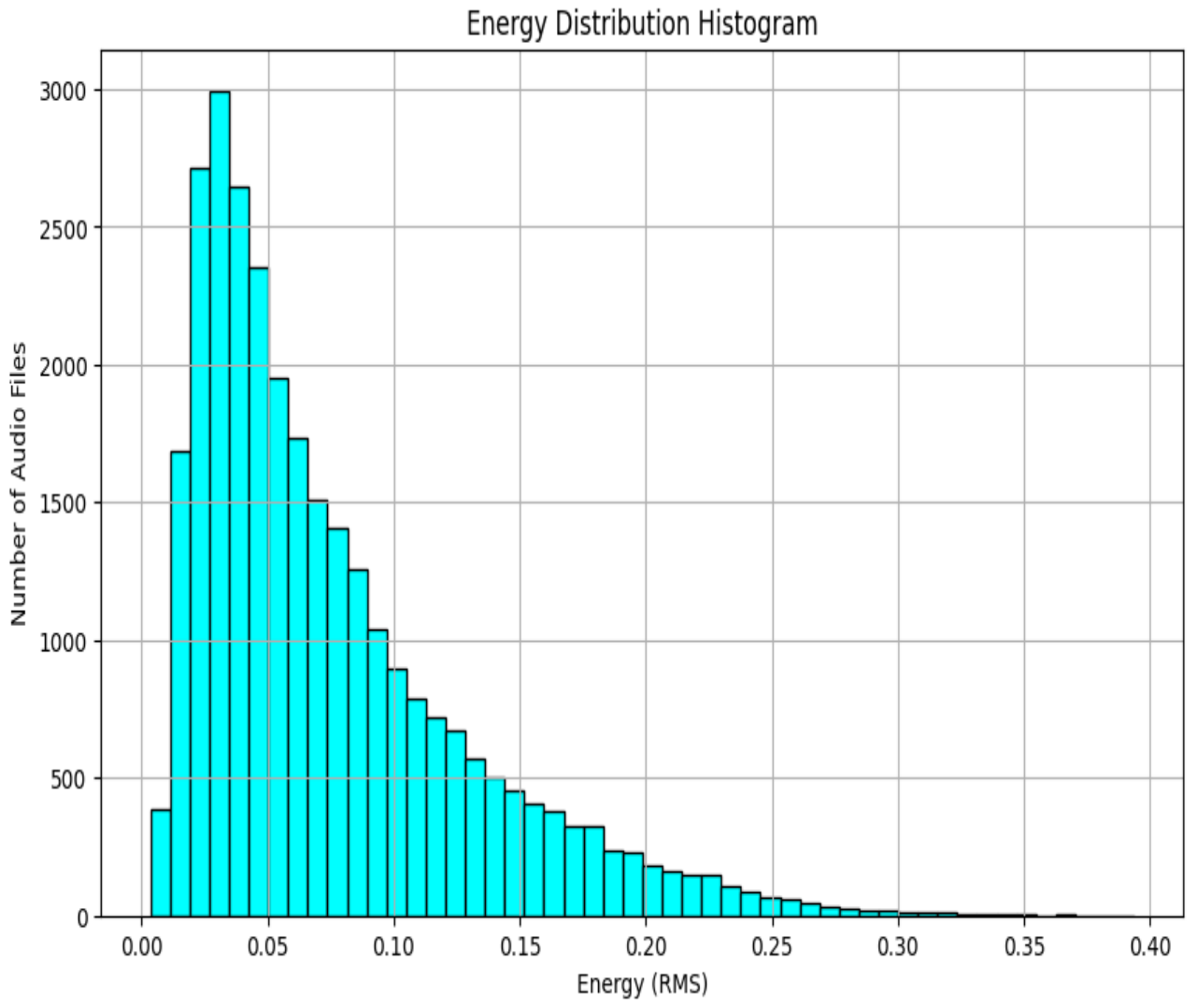

Fig.4. Histogram of Audio File Energy Levels Based on RMS Values for dataset TAU-UAS dataset

TABLE I. Outputs of the ensembled network, where a1 means the average of the teachers' logits, and the multi-letter symbols in the table mean combinations of the network's outputs.

| Ens section | a1 | z1 | z2 | a1z1 | a1z2 | z1z2 | a1z1z2 |
|---|---|---|---|---|---|---|---|
| Acc(%) | 63.4 | 64.1 | 60.1 | **64.3** | 61.1 | 60.9 | 62.4 |

### B. *Performance Analysis of the Trained Model on the TAU-UAS 2022 Mobile Dataset*

On the TAU Urban Acoustic Scenes 2022 Mobile (TAU-UAS 2022 Mobile) benchmark, we employ knowledge distillation together with an ensemble network to improve robustness under matched evaluation conditions. All results reported in Table II are trained exclusively on the 25% split of the dataset, and this same split is used for every ablation and comparison to ensure a fair, apples-to-apples assessment. Under this setup, our approach surpasses the state of the art on TAU-UAS 2022 Mobile while adhering to the official evaluation protocol. Also, our model has the number of computations and parameters within the range of the DCASE competition.

Finally, we will evaluate the results of our own model based on Table III for the DCASE 2025 task. The goal of the competition this year is to increase performance on the TAU dataset, which contains audio samples recorded by various recording devices. To enable a fair comparison with existing models in the competition, we independently trained both the ensemble and student models on data from each individual device within the training set. The training dataset comprised six distinct devices. For unseen or unknown devices, we employed a student model trained on the complete set of samples from all devices. During testing, each sample was processed using the network corresponding to its device, while samples originating from devices absent in the training phase were evaluated using the global model.

TABLE II. Performance comparison on the TAU dataset using 25% of the training data.

| models | Accuracy(%) | Params | MACs |
|---|---|---|---|
| Shao_NEPUMSE [21] | 59.7 | 100k | **16M** |
| Han_SJTUTHU [22] | 59.1 | **60k** | 30M |
| Park_KT [23] | 58.4 | **60k** | 26M |
| Chen_SCUT [24] | 58 | 120k | **16M** |
| **Ours** | **59.9** | **60k** | 30M |

Table III. Performance comparison on the TAU dataset using the devices' sample information. The results are reported based on the overall dataset and the development challenge.

| models | Accuracy(%) | Params | MACs |
|---|---|---|---|
| Karasin_JKU [25] | 60.5 | **60k** | 30M |
| Tan_SNTLNTU [26] | 60.4 | 116k | **10M** |
| Li_NTU [27] | 59.3 | **60k** | 17M |
| Luo_CQUPT [28] | 59 | **60k** | 28M |
| **Ours** | **60.6** | **60k** | 30M |

## VI. Conclusion

We introduced a compact acoustic scene classification framework that unifies an efficient student network with a learned teacher–ensemble and knowledge distillation. The design emphasizes deployability—via a lightweight, quantization-ready backbone with global response normalization—while retaining the representational strength of a diverse set of teachers through two complementary fusion heads (z1 and z2). This combination enables a single small model to approximate ensemble behavior and remain robust to device and noise variability.

Evaluations on TAU Urban Acoustic Scenes 2022 Mobile confirm the effectiveness of the approach: under matched edge-oriented constraints and protocols, the distilled student achieves state-of-the-art (SOTA) performance on the TAU dataset. Ablations indicate that the gains stem from three pillars: (i) the student architecture with GRN, (ii) ensemble-guided supervision via z1/z2, and (iii) device-aware training. Together, these components narrow the gap between research-grade accuracy and practical on-device inference.

Future work will explore tighter integration of device metadata and uncertainty-aware fusion, semi-supervised adaptation to unseen recording conditions, and broader validation across additional ASC benchmarks and real-world deployments.